
\input harvmac

\def\a{\alpha}
\def\A{{\cal A}}

\def\b{\beta}

\def\h{\hat}
\def\k{\kappa}
\def\l{\lambda}
\def\M{{\cal M}}
\def\p{\partial}
\def\t{\tilde}
\def\({\left (}
\def\){\right )}
\def\[{\left [}
\def\]{\right ]}
\gdef\journal#1, #2, #3, 19#4#5{
{\sl #1~}{\bf #2}, #3 (19#4#5)}

\lref\btz{M. Banados, C. Teitelboim, and J. Zanelli,
\journal Phys. Rev. Lett., 69, 1849, 1992;
M. Banados, M. Henneaux, C. Teitelboim, and J. Zanelli, ``Geometry of
the 2+1 Black Hole", to appear in Phys. Rev. D, gr-qc/9302012.}

\lref\kikkawa{ K.~Kikkawa and M.~Yamasaki,
\journal Phys. Lett., B149, 357, 1984;
N.~Sakai and I.~Senda,
\journal Prog. Theor. Phys., 75, 692, 1986;
V.~Nair, A.~Shapere, A.~Strominger and F.~Wilczek,
\journal Nucl. Phys., B287, 402, 1987.}

\lref\ho{For a review, see G. Horowitz, ``The Dark Side of String Theory:
Black Holes and Black Strings", to appear in the 1992 Trieste Spring School on
String Theory and Quantum Gravity.}

\lref\ka{N. Kaloper, ``Miens of the Three-Dimensional Black Hole",
Alberta-THY-8-93, hep-th/9303007; A. Ali and A. Kumar, ``$O(d,d)$
Tranformations
and 3D Black Hole", IP-BBSR-92-80, hep-th/9303032.}

\lref\kj{G. Kang and T. Jacobson, ``Conformal Invariance of Black Hole
Temperature", NSF-ITP-93-91, gr-qc/9307002.}

\lref\hw{G. Horowitz and D. Welch,
\journal Phys. Rev. Lett., 71, 328, 1993.}

\lref\gw{G. Gibbons and D. Wiltshire, \journal Ann. Phys., 167, 201, 1986;
({\bf E}), {\it ibid.} {\bf 176} (1987) 393.}

\lref\ms{J. Maharana and J. Schwarz,  \journal Nucl. Phys., B390, 3, 1993.}

\lref\hoho{ J.~Horne and G.~Horowitz,
\journal Nucl. Phys., B368, 444, 1992.}

\lref\hohorot{J.~Horne and G.~Horowitz,
\journal Phys. Rev., D46, 1340, 1992.}

\lref\jm{T. Jacobson and R. Myers, \journal Phys. Rev. Lett., 70, 3684, 1993;
M. Visser, \journal Phys. Rev. D, 48, 583, 1993; R. Wald, ``Black Hole
Entropy is Noether Charge", EFI-93-42, gr-qc/9307038.}

\lref\bch{J. Bardeen, B. Carter, and S. Hawking, \journal Commun. Math. Phys.,
31, 161, 1973.}

\lref\haw{S. Hawking, \journal  Commun. Math. Phys., 25, 152, 1972.}

\lref\rw{I. Racz and R. Wald, \journal Class. Quantum Grav., 9, 2643, 1992.}

\lref\witten{E. Witten, \journal
Phys. Rev., D44, 314, 1991.}

\lref\waldb{R. Wald, General Relativity, Univ. of Chicago Press, 1984.}

\lref\kw{B. Kay and R. Wald,
\journal Phys. Rep., 207, 49, 1991.}

\lref\hs{ G.~Horowitz and A.~Strominger,
\journal Nucl.~Phys., B360, 197, 1991.}

\lref\gr{A. Giveon and M. Ro\v cek,
\journal Nucl. Phys., B380, 128, 1992.}

\lref\sen{A. Sen, ``Macroscopic Charged Heterotic String", Tata preprint
TIFR-TH-92-29, hep-th/9206016.}

\lref\hhs{J. Horne, G. Horowitz, and A. Steif,
\journal Phys. Rev. Lett., 68,
568, 1992. This paper considered the mass per unit length, but it is easy to
verify that when the change in the dilaton is included, the total mass is
duality invariant.}

\lref\rv{M. Ro\v cek and E. Verlinde,
\journal Nucl. Phys., B373, 630, 1992.}

\lref\buscher{T. Buscher,
\journal Phys. Lett., B201, 466, 1988;
\journal Phys.
Lett., B194, 59, 1987.}

\Title{\vbox{\baselineskip12pt\hbox{NSF-ITP-93-89}
\hbox{hep-th/9308077}}}
{\vbox{
\centerline{Duality Invariance of the}
\centerline{Hawking Temperature and Entropy}}}

\centerline{{Gary T. Horowitz}\footnote{$^*$}
{On leave from the Physics Department, University of California, Santa Barbara,
CA.}}
\vskip.1in
\centerline{\sl Institute for Theoretical Physics}
\centerline{\sl University of California}
\centerline{\sl Santa Barbara, CA 93106-9530}
\centerline{\sl gary@cosmic.physics.ucsb.edu}
\vskip .1in
\centerline{Dean L. Welch}
\centerline{\sl Department of Physics}
\centerline{\sl University of California}
\centerline{\sl Santa Barbara, CA 93106-9530}
\centerline{\sl dean@cosmic.physics.ucsb.edu}

\bigskip
\centerline{\bf Abstract}
We consider solutions to low energy string theory which have a horizon
and a spacelike symmetry. Each of these solutions has a geometrically
different dual
description. We show that the dual solution has a horizon with exactly
the same Hawking temperature (surface gravity) and entropy (area)
as the original solution.

\Date{8/93}

One of the fundamental differences between theories based on
point particles and those based on strings is  that the latter
have a new type of symmetry called
duality. Geometrically different spacetimes can be shown to be
physically equivalent solutions to string theory. The simplest example
consists of flat spacetime with one direction compactified to form a circle
of radius $R$. This solution turns out to be equivalent to the one with
radius $1/R$ \kikkawa. More generally, it has been shown that any solution with
a spacelike symmetry with compact orbits has an equivalent dual
description \buscher\rv.
(If there are $d$ symmetry directions, there are many equivalent
descriptions which are related to each other by  $O(d,d,Z)$ transformations
\gr.)

Dual solutions can in general
have quite different geometric properties.
For example, solutions with curvature
singularities can be equivalent to ones without. In a sense,
certain aspects of the
spacetime geometry are pure gauge. In this paper, we examine the
effect of duality on Hawking evaporation. Since the thermodynamic properties
of spacetimes with horizons, such as their temperature and entropy, are
related to geometric properties of the solution, it is
not at all obvious that dual solutions will describe Hawking evaporation in
the same manor.  Nevertheless,
we will show that this is the case: If a solution to low energy string
theory has a horizon and a spacelike symmetry, then the dual
solution also has a horizon with exactly the same temperature. (More
precisely, we prove that this is the case for solutions with a certain
type of horizon called a bifurcate Killing horizon, which will be defined
below.) Furthermore, the area of the horizon is invariant under duality.
This is rather surprising in light of the fact that duality essentially
replaces the radius $R$ with $1/R$. However, this description is appropriate
for a metric which does not have the standard Einstein action. It is the area
in the conformally related
Einstein metric which is associated with the entropy and turns out to
be duality invariant.
Combining these results with
the fact that the spectrum of the string is the same
in dual solutions, one concludes that evaporation will occur at the same rate.

Our results clearly apply to solutions with both a horizon and a spacelike
symmetry. Of most interest are the black string solutions \ho, where the
symmetry is an asymptotic translation. In this case, the original solution
and its dual are both asymptotically flat, and the mass
is duality invariant \hhs.
The invariance of the Hawking temperature and entropy
under duality also applies
to higher dimensional extended objects surrounded by an event horizon,
i.e., black p-branes \hs.
We will also consider the three dimensional black hole \btz\hw\ka,
where the spacelike
symmetry
is a rotation. This solution is dual
to a three dimensional black string \hw. This example is somewhat unusual
since the black hole is not asymptotically flat, but rather asymptotically
anti-de Sitter. As a result, there is some ambiguity in the definition of
the Hawking temperature. Unfortunately, our results do not apply to four
(or higher) dimensional black holes since the rotational Killing field
now vanishes on the axis, which causes the horizon in the dual spacetime
to become singular.

The low energy action that arises in string theory  is
\eqn\saction{ S = \int {\rm d}^D x \sqrt{- g} e^{-2\phi}\[
		       R + 4 (\nabla \phi)^2 - {1\over 12} H^2 \]\,,}
where $\phi$ is the dilaton and $H_{\mu\nu\rho}$ is the three form. Since
$H$ is closed,
one can define a two form
potential $B_{\mu\nu}$ by $H=dB$.
Consider an asymptotically flat $D$ dimensional solution
with two commuting Killing vectors $\xi$ and $\eta$, where
$\xi$ is timelike asymptotically and $\eta$ is spacelike asymptotically
with
compact orbits. We will assume that the solution has a bifurcate Killing
horizon \kw. Roughly speaking, this means that the horizon is qualitatively
like Schwarzschild, with both a past and future component.
The precise requirement is simply that there is a constant $\Omega$
such that $ \chi \equiv
\xi + \Omega
\eta =0$ on a compact $D-2$ dimensional spacelike surface $\Sigma$.
The bifurcate Killing horizon is defined to be the future and past light
cone of $\Sigma$.
One can show  that $\chi$ is normal to this light cone, i.e.,
it is null and tangent to this surface. $\Omega$ is called the
velocity of the horizon.
Our final assumption is that $\eta^\mu \eta_\mu \ne 0$ on the horizon.

Bifurcate Killing horizons are, in fact, quite common.  The event horizons in
all  known solutions to low energy string theory are of this type,
and this is probably true for all solutions as well. The reason is
the following. Since the dilaton is invariant under all the symmetries of
the solution, and a horizon is conformally invariant, one can consider
the rescaled metric with the standard Einstein action. This has the advantage
that the matter fields now satisfy the dominant
energy condition. Under certain assumptions, Hawking has shown \haw\ that the
event horizon of every stationary black hole must be a Killing horizon,
i.e., a null hypersurface to which a Killing field is normal. (This is
trivially true for static solutions.) Since the
dominant energy condition is satisfied, the surface gravity $\kappa$ of every
Killing horizon must be constant \bch. It then follows \rw\
that if $\kappa \ne 0 $,
the horizon is in fact a bifurcate Killing horizon. Since most of these
results are local, they should apply to black strings as well as black holes.

Let us introduce coordinates so that $\eta = \p/\p x$ and $\xi = \p/\p t$.
We now derive two properties which will be
needed for the discussion of duality. We will show that everywhere on a
bifurcate Killing horizon,
\eqn\prop{  A_t \equiv {g_{tx}\over g_{xx}} = - \Omega, \qquad {\rm and}
\qquad B_{tx} = 0}
On  the surface $\Sigma$,
$\xi = -\Omega \eta$. Taking the inner product with $\eta$ yields
$g_{t x} = -\Omega g_{xx}$, which implies $A_t \equiv  g_{tx}/g_{xx} =
-\Omega$ on $\Sigma$.
By antisymmetry, $B_{tx}\equiv B_{\mu\nu}
\xi^\mu \eta^\nu = B_{\mu\nu} \chi^\mu \eta^\nu$. Since $\chi = 0 $ on
$\Sigma$, we conclude that $B_{tx} = 0 $ on this surface. Thus the two
properties \prop\ hold on $\Sigma$. Now consider $A_t$ and
$B_{tx}$ at an arbitrary point on the horizon.
Since the null
generators of the horizon are an isometry, $A_t$ and $B_{tx}$ must be constant
along these null curves. But one can get arbitrarily close to $\Sigma$ by
this isometry. It then follows by continuity that, at any point on the
horizon, $A_t$ and $B_{tx}$ must be equal to their values
on $\Sigma$. It should be noted that for spherically symmetric
solutions, if $r$ is the area coordinate, $H_{txr}$
need not vanish on the horizon, since the vector $\p/\p r$ diverges at
the bifurcation surface and compensates for the vanishing of $\chi$.

Since we are considering solutions having a spacelike symmetry
with compact orbits, there is
a dual solution which is physically equivalent in string theory \rv .
The duality transformation takes a simple form if one writes the
string metric as follows
\eqn\newform{ds^2 = g_{xx} (dx+ A_\a dy^\a)^2 +\bar g_{\a\b}dy^\a dy^\b }
where $\a$ and $\b$ run over all coordinates except $x$, and
all metric components are independent of $x$.
Under duality, $\bar g_{\a\b}$ is invariant, and the other fields transform as
\eqn\duality{ \t g_{xx} = 1/g_{xx}, \qquad \t A_\a = B_{x \a}, \qquad
  \t B_{x \a} = A_\a, \qquad \t B_{\a\b} = B_{\a\b} - 2A_{[\a} B_{\b] x}}
  $$ \t \phi =  \phi - {1\over 2} \ln g_{xx}$$
The statement that the vector $\chi$ vanishes on the surface $\Sigma$
is independent of the metric and other fields on the spacetime. Thus the
dual solution will have a bifurcate Killing horizon provided that
the future and past light cones of $\Sigma$ are nonsingular.
But this follows
immediately from the fact that
$B_{\mu\nu}$ is smooth and $g_{xx}$ is nonzero in a neighborhood of
the horizon in the original spacetime. Thus duality preserves
a bifurcate
Killing horizon. Since it also preserves the two translational symmetries
it might appear that the velocity $\Omega$ is duality invariant. However
this is not the case. The problem is that $\Omega$ is defined by the
condition $\chi = \xi + \Omega \eta = 0$ on $\Sigma$, where
$\xi$ is chosen to be the
time translation symmetry orthogonal to $\eta$ at infinity. But the
duality transformation \duality\ can change the metric asymptotically so
that $\xi$ is no longer orthogonal to $\eta$. ($\eta$ is fixed to be
the symmetry with compact orbits and is invariant under duality.) When
$\chi$ is expressed in terms of the vector which is orthogonal to $\eta$,
the  velocity may be different.

The duality transformation \duality\ has some gauge freedom. One can
add the gradient of a function to $A_\a$, and the curl of a vector to
$B_{\a\b}$. Adding $\nabla_\a f$ to
$A_\a$ corresponds to the coordinate transformation $x=\h x + f$.
One must sometimes use this freedom to insure that the dual fields are
smooth. For example, we have seen that
$A_t = -\Omega$ on the horizon. Using \duality\ we would find that the
dual solution has $\t B_{xt} = -\Omega$ on the horizon which contradicts \prop.
To avoid this, we must first define a new coordinate
\eqn\xhat{\h x = x - \Omega t}
In terms of the coordinates $(\hat x, t, y^i)$
the vector $\chi$ is simply
$\p /\p t$ and $\h A_t \equiv A_t +\Omega =0$ on the horizon. In the dual
solution\foot{Recall that a coordinate basis vector is defined
to be tangent to the
curves obtained by holding the other coordinates fixed. Thus changing $x$
to $\h x = x - \Omega t$ changes $\p/\p t$ but
$\p/\p x = \p/\p \h x$. In particular, $g_{\h x \h x} = g_{xx}$ and
dualizing with respect to $\h x$ is equivalent to
dualizing with respect to $x$.} $\t B_{\h x t}$ will then vanish on the
horizon.

We now consider
the temperature of the horizon, which is equal to the surface gravity divided
by $2\pi$. Set $\l = \chi^\mu \chi_\mu$. Then the surface gravity $\kappa$
is defined by\foot{Since
$\l =0$ on the horizon, its gradient must be normal to the surface and
hence some multiple of $\chi$. In writing the index $\a$ we are using the
fact that $\lambda$ is independent of $x$ and $\chi_\mu \eta^\mu = 0$ on
the horizon.}
\eqn\sg{\nabla_\a\lambda = -2 \kappa \chi_\a}
In terms of the coordinates $(\h x, t, y^i)$ we have just introduced,
the vector $\chi$ is simply
$\p /\p t$ so \sg\ reduces to $\nabla_\a g_{tt} = -2 \kappa g_{\a t}$.
Using the form of the metric in \newform\ and the definition of $\h x$ \xhat,
we have
$g_{tt} = g_{xx} \h A_t^2 + \bar g_{tt}$ and $ g_{\a t} = g_{xx} A_\a \h A_t +
\bar g_{\a t}$.
But $\h A_t$ vanishes at the horizon and hence the definition of the
surface gravity depends only on the duality invariant part of the metric,
$\bar g_{\a\b}$.
We conclude that the Hawking temperature is duality
invariant. (This argument is valid even though $t$ is not well behaved
at the horizon, since replacing $t$ by e.g.  an ingoing null coordinate
$v$ does not change the vector $\p/\p t = \p/\p v$.)

To discuss the entropy, it is convenient to first rescale the metric in
\saction\
so that one has the standard Einstein action. In $D$ dimensions, this
is accomplished with
the conformal factor $e^{-4\phi/(D-2)}$.
Thus the Einstein metric can be expressed
\eqn\emetric{ds^2_E = e^{-4\phi/(D-2)} [g_{xx} (dx+ A_\a dy^\a)^2
		+\bar g_{\a\b}dy^\a dy^\b] }
It is worth noting that the surface gravity computed from the Einstein
metric is identical to that computed from the (string) metric we have been
using previously. This is because $\kappa$ is invariant under arbitrary
conformal rescaling as long as $\chi^\mu$ is kept fixed \kj. Since
we now have the standard Einstein action, the entropy should be proportional
to the area\foot{By ``area" we mean the $D-2$ dimensional volume.} of the
horizon.
The metric induced on a cross
section of the horizon  takes the form \emetric\  where the indices $\a,\b$
run over the $D-3$ angular variables.  The area is thus
\eqn\area{ \A_H = \int e^{-2\phi} \sqrt{g_{xx}} \sqrt {\bar g} }
But from \duality, $e^{-2\phi} \sqrt{g_{xx}}$ is duality invariant. So
the area is also duality invariant. Notice that it is the full area of the
horizon  and not the ``area at fixed $x$" which is invariant. In fact,
it is crucial for this argument that
the horizon cross section is $D-2$ dimensional.
If we restrict the $y^\a$ to an $m$ dimensional subspace, it would have
area
\eqn\narea{ \A_m = \int e^{-2(m+1)\phi/(D-2)} \sqrt{g_{xx}} \sqrt {\bar g}}
which is not duality invariant. On the other hand, the only properties
of the horizon that we have used are its dimension and invariance under
translation of $x$: The area of every $D-2$ dimensional, $x$ invariant
surface is unchanged under duality.

Another approach to showing that the thermodynamic properties should be duality
invariant is to use the analogy with Kaluza-Klein theory \gw\ and
pass to a $D-1$ dimensional description. Since we are assuming all fields are
independent of $x$, the original action \saction\ can be written
\eqn\reduced{S= \int d^{D-1} y e^{-2\phi} \sqrt{g_{xx}}\sqrt{ -\bar g}
    [R(\bar g) + \cdots]}
where the dots denote terms involving the ``matter fields" $A_\a, g_{xx}$
as well as $\phi$ and $B_{\mu\nu}$. (Their specific form can be found
e.g. in \ms\ but will not be important here.)
Since $e^{-2\phi} \sqrt{g_{xx}}$ is duality invariant, the rescaled
$D-1$ dimensional Einstein metric is also duality invariant. In other
words, in this formulation
the duality transformation acts only on the matter fields and
all geometric properties of the solution  are duality invariant.
The difficulty is that this approach obscures the connection with
$D$ dimensional quantities. Since we have not assumed that the symmetry
direction has a small ``radius", these are the quantities of most physical
interest. For example, the area of any $m$ dimensional
surface computed with the reduced metric is duality invariant, but is
not simply related to the area of the corresponding $m+1$ dimensional
surface computed with the $D$ dimensional metric. In addition,
certain properties such as the fact that the velocity of the horizon
can change under duality, are harder to see in the reduced formulation.

As one example,
we consider a black string solution in dimension $D >4$.
The simplest black string is the product of the $D-1$ dimensional
Schwarzschild solution and $S^1$.
The boosted form of this solution, with velocity $v=\tanh \a$, is
\eqn\bhdbs{ds^2=-\(1-{r_o^n\cosh ^2{\a} \over r^n}\)dt^2+\(1+{r_o^n\sinh ^2{\a}
\over
r^n}\)dx^2-{2 r_o^n\sinh{\a} \cosh{\a} \over r^n}dxdt}
$$+\(1-{r_o^n\over
r^n}\)^{-1}dr^2 + r^2d\Omega ^2_{n+1} $$
$$B_{\mu \nu}=0, \qquad \phi=0  $$
where $r_o$ is a constant related to the original Schwarzschild mass,
$n=D-4$ and $x$ is identified with $x+1$.
This solution can be put into the form \newform\  with
\eqn\pif{ g_{xx} = \(1+{r_o^n\sinh ^2{\a}\over r^n}\)\qquad A_t=-{r_o^n\sinh \a
  \cosh \a  \over r^n + r_o^n \sinh^2 \a} }
$$ \bar g_{\a\b} dy^\a dy^\b = -\({r^n - r_o^n\over r^n+r_o^n\sinh^2 \a} \)dt^2
+\(1-{r_o^n\over r^n}\)^{-1}dr^2+r^2d\Omega ^2_{n+1}$$
The surface $r = r_o $ is a bifurcate Killing horizon. The
velocity of the horizon is $\Omega = - A_t(r = r_o) = \tanh \a$,
which is just the velocity of the boost, as expected. Introducing the new
coordinate $\h x = x - \Omega t$ we find that $\p/\p t$ is tangent to the
horizon. The surface gravity \sg\ must be calculated in coordinates which
are good on the horizon. The result is
$\k=n(2r_o\cosh{\a})^{-1}$. The area of the horizon is $c_{n+1}
r_o^{n+1} \cosh \a$ where $c_{n+1}$ is the area of a unit $n+1$ sphere.

Dualizing this solution  gives the charged black string in $D$ dimensions \hhs
\eqn\dbhdbs{d\t s^2=-{r^n-r_o^n\over r^n+r_o^n\sinh ^2{\a}
}dt^2+ \(1+{r_o^n\sinh ^2\a \over r^n}\)^{-1}d\h x^2+ \(1-{r_o^n\over
r^n}\)^{-1} dr^2+r^2d\Omega ^2_{n+1}}
$$ \t B_{\h xt}=\tanh \a \({r^n - r_o^n\over r^n + r_o^n\sinh^2\a} \),
\qquad \t \phi = -{1\over 2}\ln{\(1+{r_o^n\sinh ^2\a \over r^n}\)} $$
The surface $r = r_o$ is still a bifurcate Killing horizon,
but the velocity $\Omega$ is now zero.
The velocity has changed since, in terms of the new coordinate $\h x$,
$\h A_t$ does not vanish at infinity in the solution \pif.
Since $B_{\mu\nu}=0$, the duality
transformation changes the metric
at infinity.
The surface gravity is the same as the boosted string
$\k=n(2r_o\cosh{\a})^{-1}.$ To compare the entropy, we must first obtain
the Einstein metric by multiplying \dbhdbs\ by $e^{-4\t\phi/D-2} =
(1+{r_o^n\sinh ^2\a \over r^n})^{2/(n+2)}$.
The area of the horizon is then $ c_{n+1}r_o^{n+1} \cosh \a$
which agrees with the dual solution.

As another example, we consider the three dimensional black hole solution.
The metric
is \btz
\eqn\bh{ds^2=-\({r^2\over l^2} - M\)dt^2+\({r^2\over
l^2}-M+{J^2\over 4r^2}\)^{-1}dr^2 +r^2d\theta^2-Jdtd\theta}
where
$M$ and $J$ are the  mass and angular momentum of the black hole.
This solution can be obtained by identifying points of three dimensional
anti-de Sitter space.
The curvature is  thus constant and depends only on $l,$
$R_{\mu \nu} =-{2\over
l^2} g_{\mu \nu}.$ The solution has two bifurcate Killing
horizons at $r= r_\pm$ where
$r_\pm^2 = {Ml^2 \over 2} (1 \pm \sqrt{ 1-{J^2 \over M^2 l^2}}) $.

Unlike the case
of asymptotically flat solutions, there is no preferred normalization
of the timelike Killing vector. This is important since the surface gravity
\sg\ depends on this normalization. If one sets $\xi = \p/\p t$, then the
velocity of the horizon is $\Omega = J/2r_+^2$ and the
vector tangent to the horizon is $\chi = \p/\p t
+ \Omega \p/\p \theta$.
The surface gravity of the black hole is then
\eqn\temp{ \kappa = {r_+^2 - r_-^2 \over l^2 r_+}}
This has the interesting property that it decreases as the
mass decreases. It is also reduced by the presence of angular momentum.
It is important to note that $T=\kappa/2\pi$
 is not the temperature seen by observers
at infinity. This is because the black hole \bh\ is
asymptotically anti-de Sitter. For this metric, there
is an infinite redshift
between any finite radius and infinity. So the temperature at infinity
always vanishes.
Of course, just because no energy reaches infinity, does not mean that the
black hole cannot evaporate. The mass of the black hole can still be converted
into thermal radiation.

In \hw\ \ka\ it was shown that the black hole metric \bh\ along with the
fields
$B_{\theta t}=r^2/l, \phi =0,$ is a solution to string theory. To insure that
$B_{\mu\nu}$ is regular on the horizon, we must perform a gauge transformation
so that $B_{\theta t}=(r^2-r^2_+)/l$. To insure that the dual antisymmetric
tensor $\t B_{\mu\nu}$ is also regular we introduce the shifted coordinate
$\h \theta = \theta - \Omega t$.
 The result of dualizing on $\h \theta$ is then
\eqn\shif{ d\t s^2=- {(r^2-r_+^2)(r_+^2-r_-^2)\over r^2 l^2}dt^2+{2\over
{r^2l}}(r^2-r_+^2)dtd\h \theta +{d \h \theta^2\over r^2}+{{r^2l^2}\over
{(r^2-r_+^2)(r^2-r_-^2)}}dr^2}
$$\t B_{\h \theta t}={J\over 2r^2_+} - {J\over 2r^2},
\qquad \t \phi = -\ln r $$
where we have used the fact that $M=(r_+^2+r_-^2)/l^2$ and $J=2r_+r_-/l$.
The dual solution can be put in the form of a black string \hoho
\eqn\bs{ d\t s^2=-\(1-{\M\over {\h r}}\)d\h t^2 + \(1-{Q^2\over
{\M \h r}}\)d\h x^2
+ \(1-{Q^2\over {\M \h r}}\)^{-1}\(1-{\M\over {\h r}}\)^{-1} {{l^2d\h r^2}\over
{4\h r^2}}}
$$ \t B_{\h x\h t}={Q\over {\h r}}- {Q\over \M}, \qquad \t \phi =-{1\over 2}
\ln{\h r l} $$
by making the coordinate transformation $t=l(r_+^2-r_-^2)^{-1/2}(\h t+\h x),
\h \theta=(r_+^2-r_-^2)^{1/2} \h x, r^2=l\h r,$ and by the identification of
parameters $\M=r_+^2/l, Q=-J/2.$ This is simpler than the
transformation needed if the shift in $B_{\theta t}$ is not made \hw .

The black string \bs\ has a bifurcate Killing horizon at $\h r = \M$
which corresponds to $r=r_+$, the same location as the black hole horizon \bh.
The surface gravity of the black string is
$\kappa = {1\over l} (1-{Q^2 \over \M^2})^{1/2}$ which corresponds to
\eqn\moresg{ \kappa = {(r^2_+ - r_-^2)^{1/2} \over l r_+} }
This does not agree with \temp. In particular, when $Q=0$, $r_- = 0$ and
$\kappa$ is
independent of the mass.
The reason for the difference can be traced back to the
ambiguity in the normalization of the timelike Killing field for the
three dimensional black hole. The black string is asymptotically flat
and so its surface gravity is unambiguous. However, in order to obtain
the standard form of the metric \bs\ from the dual of the black hole \shif,
we had to rescale the time coordinate. Since the Killing field
has been taken to be $\p/\p t$,  this results in a change
in the surface gravity. In other words, the unambiguous timelike Killing
field in the black string solution \bs\ does not correspond, under
duality, to $\p/\p t$ in \bh, but rather some multiple of this.

The area of the horizon is defined unambiguously in both the black hole
solution and its dual.  We now show
that these
two areas are equal, as required by our general argument. The area (length)
of the
horizon of the three dimensional black hole \bh\ is clearly $2\pi r_+$.
To compute the area of the horizon for the black string \bs, we first
multiply by $e^{-4\t\phi}= \h r^2 l^2$ to obtain the Einstein metric.
The area is then $\A = \int \h r l (1-{Q^2 \over \M \h r})^{1/2} d\h x$.
But the horizon is at $\h r = \M$ and from the relation between
$\h x$ and $\h \theta$ we have $\int d\h x = 2\pi /(r_+^2 - r_-^2)^{1/2}$.
Therefore
\eqn\bsarea{ \A= 2\pi l \({\M^2 - Q^2 \over r_+^2 - r_-^2 }\)^{1/2}
      = 2 \pi r_+}
where we have used the above expressions for $\M$ and $Q$ in terms of $r_\pm$.
Thus even though the asymptotic structure changes under duality, the
horizon area is invariant.

To summarize, we have shown that certain thermodynamic properties of solutions
to low energy string theory with horizons are invariant under duality.
This is strong evidence that the concepts of temperature and entropy
remain physically meaningful in string theory. It is rather surprising
that the duality invariant entropy is simply proportional to the area.
For solutions to the low energy theory \saction, the entropy is indeed
given by the area. But dual solutions are
certainly not physically equivalent as solutions to {\it this} theory.
It is only when viewed in the full context of string theory
that one can establish this equivalence. But the full string equations involves
higher curvature terms, and it has recently been shown that for theories
of this type, the entropy is not simply proportional to the area \jm.
Thus, one might have expected that the duality invariant entropy would
be a complicated expression involving integrals of curvature terms over
the horizon. We have seen that this is not the case: The simplest expression
is  already invariant under the low energy duality transformation.
This is the first step toward finding the
exact expression for entropy in
string theory.
\vskip 1cm

\centerline{Acknowledgments}
\vskip .5cm
It is a pleasure to thank W. Nelson and R. Wald for discussions.
This work was supported in part by NSF Grants PHY-8904035 and
PHY-9008502.

\listrefs

\end